\documentclass[acmsmall]{acmart}

\setcopyright{acmcopyright}
\acmJournal{JETC}
\acmYear{2020} \acmVolume{1} \acmNumber{1} \acmArticle{1} \acmMonth{1} \acmPrice{15.00}\acmDOI{10.1145/3397513}

\usepackage{threeparttable}
\usepackage{tabularx}
\usepackage{booktabs}
\usepackage{caption}
\usepackage{colortbl}

\usepackage{subfigure}
\usepackage{graphicx}
\usepackage{multirow}
\usepackage{booktabs}
\usepackage{flushend}
\usepackage{textcomp}
\usepackage{tablefootnote}

\begin{document}


\title{Hardware Security in Spin-Based Computing-In-Memory: Analysis, Exploits, and Mitigation Techniques}

\author{Xueyan Wang}
\author{Jienlei Yang}
\authornote{Corresponding authors: Jianlei Yang and Weisheng Zhao. Email: jianlei@buaa.edu.cn, weisheng.zhao@buaa.edu.cn.}
\author{Yinglin Zhao}
\author{Xiaotao Jia}
\affiliation{%
  \institution{Beihang University}
  \streetaddress{No. 37 Xueyuan Road, Haidian District}
  \city{Beijing}
  \country{China}
  \postcode{100191}
}

\author{Gang Qu}
\affiliation{%
  \institution{University of Maryland, College Park}
  \state{MD}
  \country{USA}
}

\author{Weisheng Zhao}
\authornotemark[1]
\affiliation{%
  \institution{Beihang University}
  \city{Beijing}
  \country{China}
}

\thanks{
This work is supported in part by State Key Laboratory of Computer Architecture (CARCH-201917), National Natural Science Foundation of China (61602022, 61701013), State Key Laboratory of Software Development Environment (SKLSDE-2018ZX-07), National Key Technology Program of China (2017ZX01032101) and the 111 Talent Program B16001.
}
\authorsaddresses{Authors’ addresses: X. Wang, Y. Zhao, X. Jia, and W. Zhao are with the Fert Beijing Research Institute, School of Microelectronics, Beijing Advanced Innovation Center for Big Data and Brain Computing (BDBC), Beihang University, No. 37 Xueyuan Road, Haidian District, Beijing, China, 100191; J. Yang is with the School of Computer Science and Engineering, BDBC, Beihang University, Beijing, China; Gang Qu is with the Department of Electrical and Computer Engineering, A.V. Williams Building, University of Maryland, College Park, MD, USA, 20742}

\renewcommand{\shortauthors}{X. Wang et al.}

\begin{abstract}

Computing-in-memory (CIM) is proposed to alleviate the processor-memory data transfer bottleneck in traditional Von-Neumann architectures, and spintronics-based magnetic memory has demonstrated many facilitation in implementing CIM paradigm.
Since hardware security has become one of the major concerns in circuit designs, this paper, for the first time, investigates spin-based computing-in-memory (\texttt{SpinCIM}) from a security perspective. \textit{We focus on two fundamental questions:} 1) how the new \texttt{SpinCIM} computing paradigm can be exploited to enhance hardware security? 2) what security concerns has this new \texttt{SpinCIM} computing paradigm incurred?

\end{abstract}



\maketitle

\section{Introduction}
For traditional Von-Neumann architecture, computation and memory are the two most important units. The computation unit reads data from memory and performs calculations, then stores the results back into memory. Since the late 1990's, CPU speed has outperformed the speed of memory access, creating the well-known ``memory wall'' \cite{mckee2004reflections}. This becomes even worse in today's big data era, where data intensive applications need more frequent transfer of larger amount  of data between processor and memory.
A single memory \texttt{Read/Write} operation consumes two to three orders of magnitude more energy and time than data calculation.

Computing-in-memory (CIM) is one promising approach to alleviate the memory wall. The basic idea of CIM is to embed computations into memory.
The computation could be performed at the same time of memory access, eliminating the time and energy overhead of data movement between processor and memory.
In addition, due to the available high bandwidth within memory array, CIM is able to provide massive parallelism, such as the vector operations.

Among the possible CIM implementations, spintronics-based CIM (\texttt{SpinCIM}) has the following advantages.
First, spintronics-based magnetic memory has been a promising candidate for the next generation main memory because of its properties such as near-zero leakage, non-volatility, high endurance, and compatibility with the CMOS manufacturing process. In particular, prototype Spin Transfer Torque Magnetic RAM (STT-MRAM) chip demonstrations and commercial MRAM products have been available by companies such as Everspin and TSMC \cite{wang2018current}\cite{mram2018isscc}.
Second, STT-MRAM stores data with magnetic-resistances and accesses data by current-sensing scheme instead of conventional charge based store and access. This enables MRAM to provide inherent computing capabilities with only minor changes to the memory array, making it suitable for CIM implementation.

\begin{figure}[t]
\centering
\includegraphics[width = 0.75\linewidth]{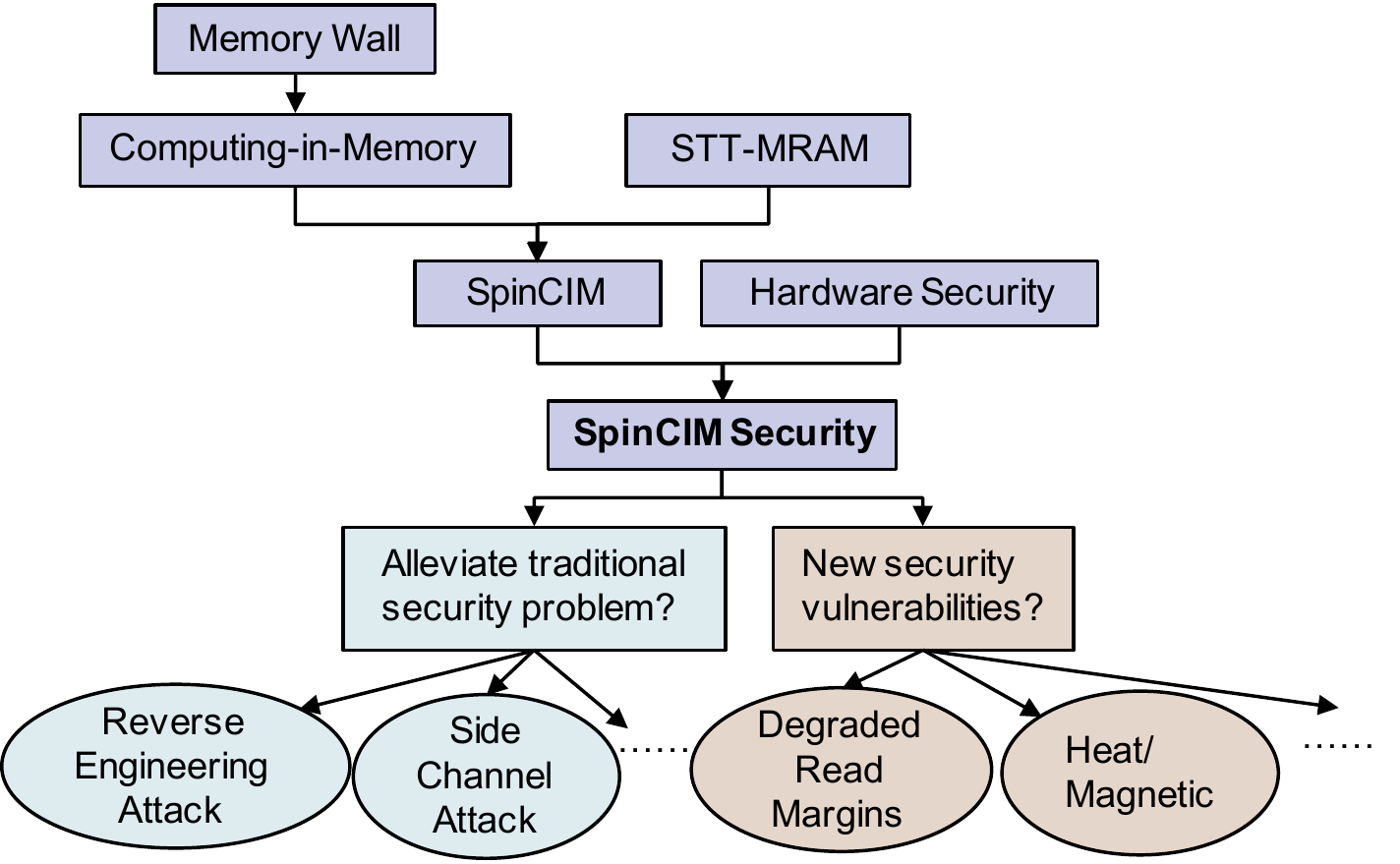}
\caption{An overview of \texttt{SpinCIM} security.}
\label{fig:overview}
\end{figure}

It is well-documented that security has been previously considered as an afterthought, with performance dominating the design requirements. As a consequence, numerous security vulnerabilities and malicious attacks (such as intentional sensitive data leakages in Meltdown and Spectre \cite{lipp2018meltdown}) are consistently being discovered.
While \texttt{SpinCIM} is newly proposed and still in exploration stage, it gives us the golden opportunity to consider security as a first class requirement at the early design stage. To the best of our knowledge, this paper is the first effort to study \texttt{SpinCIM} from the security perspective.
Previously, there are extensive research on applying the emerging spintronic device for security. For example, the stochastic writing features have been used for true random number generation and PUF design \cite{wang2016novel,iyengar2016spintronic,vatajelu2016security}. Our work is different from these in that we focus on the \texttt{SpinCIM} computing architectures, and investigate the security applications and security issues brought by this new computing paradigm.

Fig.~\ref{fig:overview} summarizes the motivation and rationale of studying \texttt{SpinCIM} security. We find that the emerging \texttt{SpinCIM} acts as a double-sided sword for hardware security. On the one hand, the new computing paradigm of \texttt{SpinCIM} facilitates the development of certain innovative hardware security solutions. In particular, we will discuss \texttt{SpinCIM}-enhanced security solutions for circuit obfuscation and side channel attack prevention. Although both topics have been heavily researched, we believe that \texttt{SpinCIM} based solutions have low performance overhead and are more effective. On the other hand, \texttt{SpinCIM} could also introduce new security vulnerabilities which can be leveraged by the attackers to launch new attacks. For example, typical hardware Trojan requires the insertion or modification of specific circuit for Trojan insertion and activation. While under \texttt{SpinCIM}, attackers are able to achieve Trojan-similar attacks by simply manipulating the thermal conditions or magnetic field, even without the need for any circuit modifications.


The rest of this paper is organized as follows: Section~\ref{sec:preliminaries} provides the necessary background knowledge on STT-MRAM and \texttt{SpinCIM}, and discusses the recent research advance in hardware security. Section~\ref{sec:cimenhance} demonstrates the enhanced security solutions for circuit obfuscation and side channel attack thwarting by \texttt{SpinCIM}. New security issues in \texttt{SpinCIM} are studied and discussed in Section~\ref{sec:cimsecurity}.
Section~\ref{sec:outlook} provides an outlook into the research for \texttt{SpinCIM} security.
Finally, Section~\ref{sec:conclu} concludes the paper.

\section{Preliminaries}\label{sec:preliminaries}

In this section, we provide a brief background introduction on STT-MRAM and \texttt{SpinCIM} as well as the current state of research on hardware security.

\begin{figure}[t]
\centering
\includegraphics[width = 0.5\linewidth]{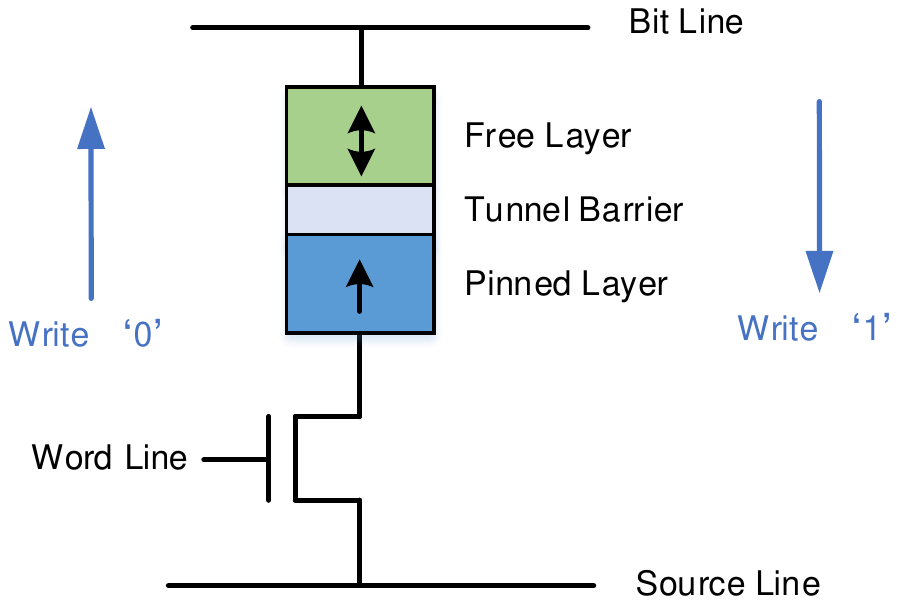}
\caption{A typical STT-MRAM bit-cell.}
\label{fig:MTJ}
\end{figure}

\subsection{Spin-Transfer Torque Magnetic RAM}

STT-MRAM is a promising next generation memory that features non-volatility, fast writing and reading, high endurance, and zero standby power \cite{DBLP:journals/pieee/KimPCKSWK15,yang2015radiation,cao2018memory,zhao2019stt,chen2009compact}.
As Fig.~\ref{fig:MTJ} shows, a typical STT-MRAM bit-cell consists of an access transistor and a Magnetic Tunnel Junction (MTJ), which is controlled by bit-line (BL), word-line (WL) and source-line (SL).
An MTJ consists of one pinned ferromagnetic layer (PL) with a fixed magnetic orientation, one free ferromagnetic layer (FL) whose magnetic orientation can be switched, and one tunneling oxide barrier between PL and FL.
The relative magnetic orientations of PL and FL can be stable in parallel (\texttt{P} state) or anti-parallel (\texttt{AP} state), corresponding to low resistance ($R_{\rm P}$) and high resistance ($R_{\rm AP}$, $R_{\rm AP}>R_{\rm P}$) of the MTJ cell, respectively. As a result, each MTJ is able to store 1-bit information.
In this paper, we assume that the low resistance state is used to represent logic `1', and the high resistance state is used to represent logic `0'.

To read out the stored information in an MTJ cell, one needs to enable WL signal, apply a voltage $V_{\rm read}$ across BL and SL, and sense the current that flows ($I_{\rm P}$ or $I_{AP}$) though the MTJ. By comparing the sense current with a reference current ($I_{\rm ref}$, $I_{\rm AP}<I_{\rm ref}<I_{P}$) the data stored in MTJ cell (logic `0' or logic `1') could be read out.
Writing operation can be performed by enabling WL, then applying appropriate voltage ($V_{\rm write}$) across BL and SL to pass a current that is greater than the critical MTJ switching current. The specific logic value that is written is dependent on the direction of the write current.

\subsection{Spin-Based Computing-In-Memory}

\begin{figure}[t]
\centering
\includegraphics[width = 0.7\linewidth]{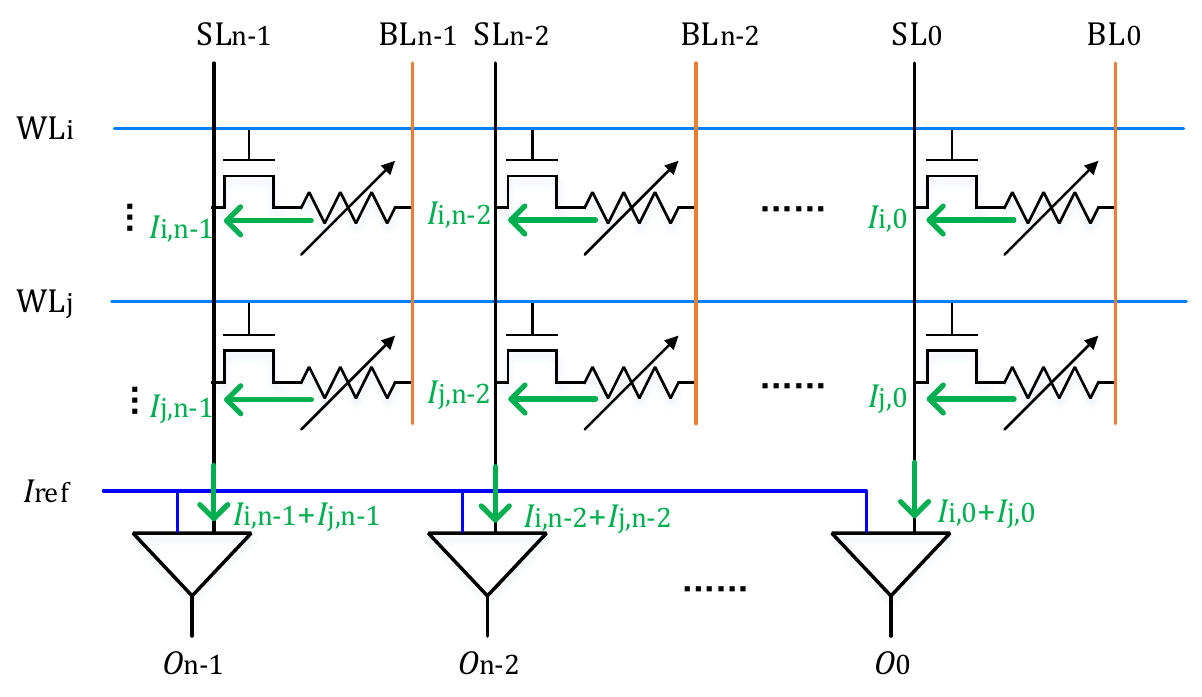}
\caption{\texttt{SpinCIM} computations in enhanced STT-MRAM array.}
\label{fig:SpinCIM}
\end{figure}

CIM efforts can be classified into two categories according to whether they target at application-specific computations \cite{DBLP:conf/isca/AhnYMC15,liu2015reno,ramasubramanian2014spindle} or general-purpose computations \cite{DBLP:journals/tvlsi/JainRRR18,DBLP:journals/cal/ChowdhuryHKZLLS18,DBLP:conf/dac/LiXZZLX16,kang2017memory,parveen2018hielm,zhao2019stt}.
{For example, ReRAM has been widely explored and used to implement the matrix-vector multiplication for neural network accelerations with the multi-bit storage property.
Compared to other resistive memory devices (such as ReRAM), STT-MRAM has higher write endurance, faster write speed, lower write energy, and limited resistance difference between the distinct resistance states of MTJ.
STT-MRAM is widely used to implement bit-wise operations for general in-memory computing paradigm.}
In this paper, we focus on such general-purpose CIM \cite{DBLP:journals/tvlsi/JainRRR18}, which can be widely used in all categories of applications.

Due to the current sensing mechanism in STT-MRAM and the fact that current can be accumulated, \texttt{SpinCIM} is able to realize logic functions conveniently.
As demonstrated in Fig.~\ref{fig:SpinCIM}, by simultaneously enabling word-line $\rm WL_i$ and $\rm WL_j$, then applying $V_{\rm read}$ across $\rm BL_k$ and $\rm SL_k$ ($\rm k \in [0,n-1]$), the current that feeds into the $\rm k$-th sense amplifier (SA) is a summation of the currents flowing through $\rm MTJ_{i,k}$ and $\rm MTJ_{j,k}$, namely $I_{\rm i,k}+I_{\rm j,k}$.
With different reference sensing current, the sense amplifier will have different outputs under given input patterns, thus different logic functions of the enabled word line can be directly implemented.

\begin{figure}[ht]
\centering
\includegraphics[width = 0.58\linewidth]{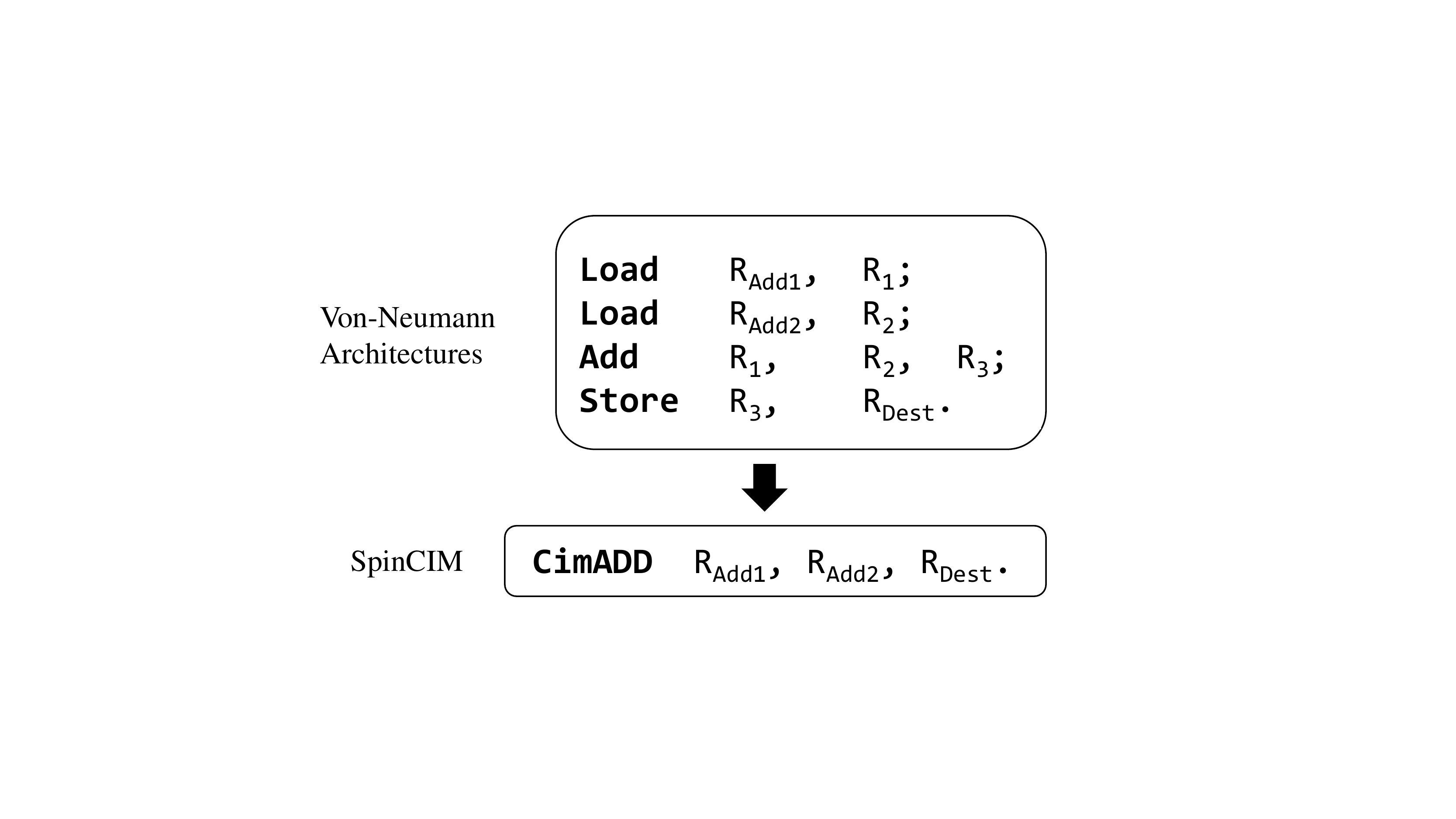}
\caption{An example of \texttt{CimADD} operation in extended ISA.}
\label{fig:ISAExample}
\end{figure}

In \texttt{SpinCIM} computing paradigm, the core bit-cell and array structure of STT-MRAM remain unchanged. One only needs to make insignificant modifications to the peripheral circuitry (such as sensing circuitry to generate required sensing current) of the memory. Therefore, the impact of introduced \texttt{SpinCIM} on density and efficiency of memory arrays is negligible \cite{DBLP:journals/tvlsi/JainRRR18}.

In the architecture level, to invoke the different types of operations that can be performed by \texttt{SpinCIM} in the enhanced STT-MRAM array, the instruction set architecture (ISA) of the processor needs to be extended.
For \texttt{SpinCIM} computations, the operation type, address of operation data, and address of result are sent to memory, then the operations can be completed within the memory. The overall computation needs only one memory access.
As a comparison, traditional computations need to fetch operation data from memory to processor, perform computations in the processor, then write the result back to memory, which involve $M+1$ ($M$ is the number of operation data that needs to fetch from memory) number of memory access.
Take the \texttt{Add} operation as an example, as shown in Fig.~\ref{fig:ISAExample}, to add two data that are stored in the main memory, the typical four instructions can be replaced by one \texttt{CimADD} instruction in \texttt{SpinCIM}. The number of memory operations is reduced from three to one, which is obviously more time and energy efficient.

\subsection{Hardware Security}

As the root of software, system, and network security, in recent years, hardware security has become a hot topic and attracted attention from both industry and academia \cite{rostami2014primer,tehranipoor2010survey}.
The continually increasing design complexity and cost have led to the globalization of integrated circuit (IC) design and fabrication, where counterfeits may exist in all phases of the supply chain and bring serious security concerns.
Design intellectual property (IP) infringement, reverse engineering attacks, hardware Trojans, side channel attacks, and others have caused both security concerns and economic loss in semiconductor industry \cite{tehranipoor2011introduction,jin2015introduction,rostami2014primer,tehranipoor2010survey}.
To ensure the integrity and the trustworthiness of fabricated circuits,
various defensive approaches have been proposed. There are passive ones like circuit watermarking/fingerprinting techniques, proactive strategies such as circuit obfuscation and Trojan detection/prevention techniques. Security primitives such as physical unclonable function (PUF) and true random number generator (TRNG) have been proposed to provide authentication and encryption.
To reduce the performance overhead of CMOS-based security techniques, researchers have explored to utilize the unique intrinsic properties of emerging devices, {for example, in circuit obfuscation strategies, polymorphic gates have been designed with the tunable polarity of SiNW FET \cite{DBLP:conf/date/ChenHJNY16} and MRAM is used to replace SRAM to configure the functionalities of multiplexers \cite{DBLP:conf/dac/WinogradSMGH16,yang2019exploiting}. However, these approaches follow the same design methodology as the previous CMOS-based ones, thus they are still vulnerable to SAT-based attacks.
Moreover, to thwart side channel attacks, it needs to slow down the fast operations and to increase the power consumption of low-energy operations to equalize the performance of different memory operations, which brings non-trivial performance overheads \cite{DBLP:conf/dft/IyengarGRN16}. This motivates us to explore the possible solutions to these security concerns with the new \texttt{SpinCIM} computing paradigm and investigate the security vulnerabilities in \texttt{SpinCIM}. Our work in this paper is different with previous work in that we focus on the \texttt{SpinCIM} computing architectures, and investigate the security issues brought by this new computing paradigm.}

\section{SpinCIM Enhanced Hardware Security}\label{sec:cimenhance}

Because of the aforementioned properties of \texttt{SpinCIM}, it can solve some of the long standing challenges in hardware security. In the section, we demonstrate this with examples on how to apply \texttt{SpinCIM} for circuit obfuscation and prevention of side channel attacks.

\subsection{Enabling Circuit Obfuscation}

Circuit obfuscation has been proposed as one proactive countermeasure against a variety of hardware attacks, such as reverse engineering (RE) and IC/IP piracy, through hiding valuable circuit design information \cite{rajendran2012security,rajendran2013security}.
Since it was proposed in 2012, the fierce race between sharpening the spears of de-obfuscation tools and making the obfuscation shield more robust has quickly elevated the sophistication and maturity level of circuit obfuscation, making it one of the most effective countermeasures against RE-based attacks and IC/IP piracy.

Two major issues exist in current circuit obfuscation techniques.
First, the recently proposed SAT-based de-obfuscation attack poses serious threats to the effectiveness of circuit obfuscation techniques.
For the current anti-SAT approaches (such as CamoPerturb \cite{DBLP:conf/iccad/jvCamouPerturb16}, And-Tree \cite{DBLP:conf/iccad/yubei16}, Anti-SAT \cite{Yasin2016SecurityAO}), they are either vulnerable to bypass/removal attacks \cite{xu2017novel,yasin2017removal}, or can be revealed by approximate attacks \cite{shamsi2017appsat,shen2017double}.
Second, the delay, power, and area overhead of the state-of-the-art circuit obfuscation strategies are so high that they cannot be applied in commercial circuits, especially when the security requirement is high.
For example, an obfuscating units that can be either \texttt{NAND}, \texttt{NOR}, or \texttt{XOR} has $5.1\times$-$5.5\times$ higher power, $4\times$ larger area, and $1.1\times$-$1.6\times$ longer delay compared to a conventional \texttt{NAND} or \texttt{NOR} logic gate \cite{rajendran2013security}.

To meet the above challenges in CMOS-based obfuscation methods, researchers have attempted to exploit the unique properties in post-CMOS emerging devices to perform obfuscation.
For example, polymorphic gates have been designed with the tunable polarity of SiNW FET \cite{DBLP:conf/date/ChenHJNY16}, MRAM is used to replace SRAM to configure the functionalities of multiplexers \cite{DBLP:conf/dac/WinogradSMGH16,yang2019exploiting}, and spintronic devices are designed to be able to perform one of the multiple functionalities with identical layout \cite{DBLP:conf/date/PatnaikRKSR18,rangarajan2018opening,DBLP:conf/glvlsi/AlasadYF17}.
By utilizing the intrinsic features (such as tunable polarity) in these emerging devices, one can reduce the performance overhead dramatically.

\begin{figure}[t]
\centering
\includegraphics[width = 0.9\linewidth]{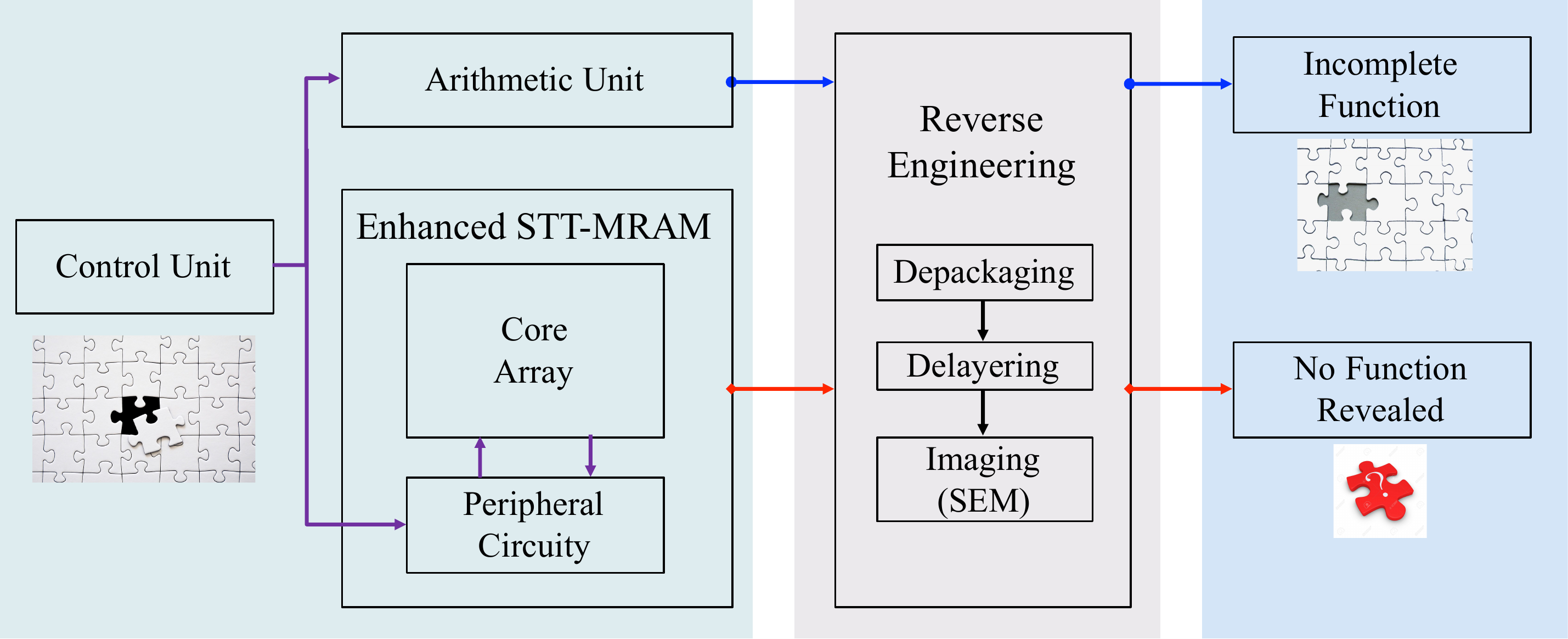}
\caption{\texttt{SpinCIM}-enabled obfuscation.}
\label{fig:SpinObfuscation}
\end{figure}

{Due to the current sensing property and the fact that current can be accumulated, spintronic is suitable to implement general-purpose computing-in-memory, and such CIM architecture facilitates obfuscating the function that is implemented in the memory array.
Therefore, SpinCIM naturally facilitates computing-in-memory and thus obfuscation.
}
Now we elaborate how \texttt{SpinCIM}-enabled obfuscation solution will have no chip area overhead and can be secure against SAT-based attacks.
Recall that in \texttt{SpinCIM} paradigm, some portion of the computation will be completed in the STT-MRAM array to reduce the data transfer between memory and CPU. The \texttt{SpinCIM} instructions, consisting of the type of the operation and the memory address of the operands, will be send to the memory array. More specifically, memory address of the operands will be sent to the address decoder unit to enable corresponding word-lines, and the type of the operation will be sent to the memory control unit. Then the control unit generates corresponding control signals to complete the computations \cite{DBLP:journals/tvlsi/JainRRR18}.
Different functions can be implemented with exactly the same hardware memory units and the same peripheral circuity of STT-MRAM memory array. It is the run-time control signals that will determine the functionality implemented by the STT-MRAM memory array (see Fig.~\ref{fig:SpinObfuscation}).
In other words, the portion of computation performed in the STT-MRAM array has been obfuscated naturally.
{
At this point, as long as the designer decides which portion to be implemented in STT-MRAM array, there is no hardware overhead for the obfuscation.
To achieve a higher level of obfuscation and protect the peripheral circuit design, the designer may obfuscate the peripheral circuity in order not to give the attacker any hint on the functions that are implemented in the STT-MRAM array.
The peripheral circuity of STT-MRAM array normally only takes a very small percentage of the whole circuit, thus the overhead, if any, will be negligible {\cite{DBLP:journals/tvlsi/JainRRR18}}.}

As the right half of Fig.~\ref{fig:SpinObfuscation} shows, an RE attacker might be able to obtain the netlist and be able to solve the portion of the computation that is implemented in the traditional CMOS technology by arithmetic unit, even it is obfuscated.
{To perform SAT-based de-obfuscation attacks, the attacker needs to get the gate-level netlist which consists of conventional logic gates and obfuscated gates, the oracle function, and the set of possible functionalities of obfuscated gates.
Even if the attacker is able to get the oracle function of the circuit, in SpinCIM-enabled obfuscation, the obfuscated functions that are implemented in the STT-MRAM array appear as identical memory array for an RE attacker, and the obfuscated peripheral circuit reveals nothing about the possible functionalities in SpinCIM.
As a result, SAT-based attacks cannot be formed,
and this solution will be secure against existing SAT-based attacks.}
Without the knowledge of this portion of the computation, the incomplete design the attacker has obtained from the traditional CMOS implementation also might become meaningless.

The key challenge for \texttt{SpinCIM}-enabled obfuscation techniques is how to split the functions into CMOS implementation and \texttt{SpinCIM} implementation and how to make them co-operate efficiently at the high level \cite{hsieh2016transparent,pattnaik2016scheduling}.
{Those computation-intensive applications with simple few control logic is suitable to be implemented by \texttt{SpinCIM}, and for specific computations, the benefits can be maximized when the operation data are from memory and the result data needs to be stored back to memory. Overall, it needs the device to algorithm level collaborations and much more related research to apply SpinCIM.}

\subsection{Thwarting Side Channel Attacks}

\begin{figure*}[t]
\centering
\subfigure[]{
\label{fig:sca1}
\includegraphics[width = 0.95\linewidth]{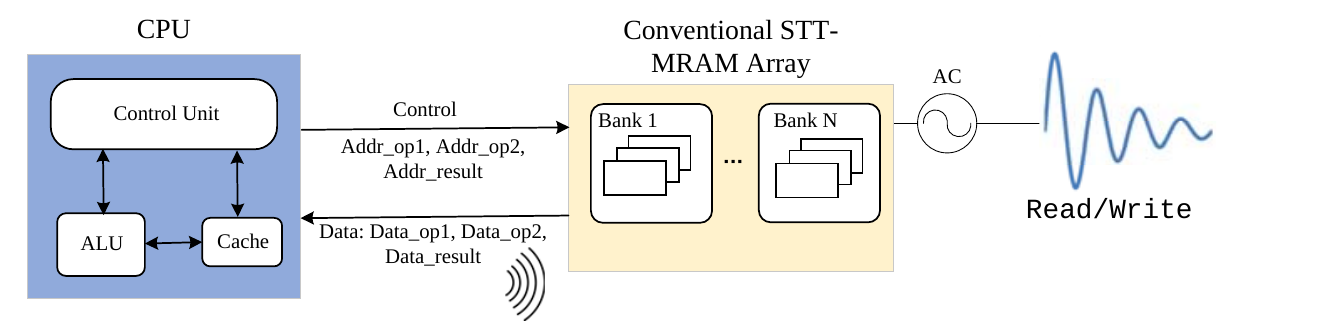}}
\subfigure[]{
\label{fig:sca2}
\includegraphics[width = 1.0\linewidth]{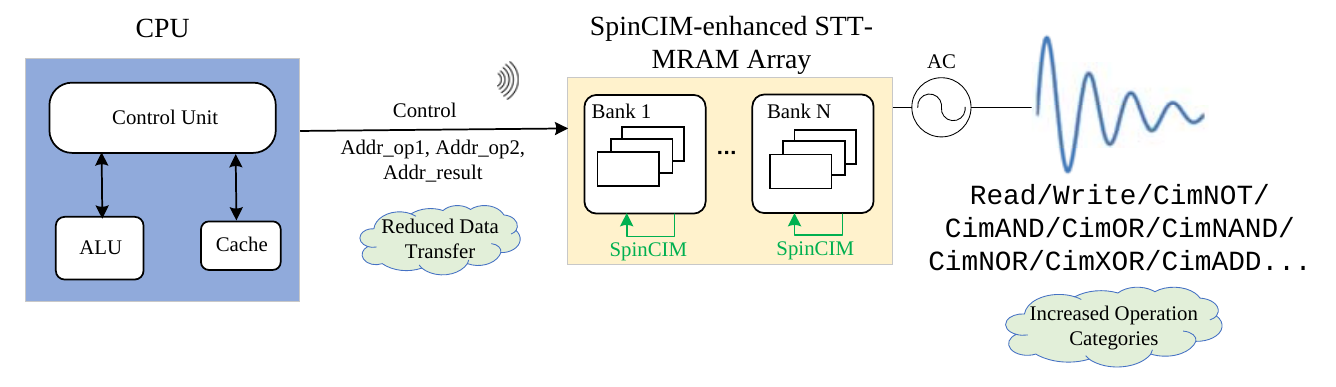}}
\caption{Reduced side channel information leakage and complex side channel analysis. (a) Traditional data transfer pattern. (b) Data transfers in \texttt{SpinCIM} computing paradigm.}
\end{figure*}

Side channel attacks aim to reveal the sensitive or secret information by analyzing the data that can be obtained during the execution of a computer system through channels such as power consumption profile and timing delay \cite{kocher1996timing}.
Since it was first reported, side channel attack has become one of the most powerful and effective attacks and poses threats to the security of computer system \cite{zhou2005side}.

In STT-MRAM array, asymmetry exists in the \texttt{Read} and \texttt{Write} operations, which facilitates statistical analysis. We have performed device and circuit simulations for memory operations.
{Specifically, at the device level, we jointly use the Brinkman model and Landau-Lifshitz-Gilbert (LLG) equation to characterize MTJ, and the key parameters for MTJ simulation are demonstrated in Table~\ref{tab:parameter}. For the circuit level simulation, a Verilog-A core block of STT-MRAM device is designed to build the circuit with a 45nm free Product Development Kit (PDK) library in Cadence.}
As demonstrated in Table~\ref{tab:StandardTimingDelay}, the latency and current of \texttt{Write} are higher than those for \texttt{Read}. Furthermore, switching to \texttt{P} state is easier than switching to \texttt{AP} state, which is known as the polarity-dependent asymmetry for \texttt{Write} latency and \texttt{Write} current. As a result, writing logic `1' and writing logic `0' incur different latency and current.
These features introduce major security vulnerabilities which have been exploited by side channel attackers who can monitor the signatures during memory \texttt{Read/Write} operations to compromise data privacy \cite{DBLP:conf/dft/IyengarGRN16,DBLP:conf/iccad/GhoshKDJ16,DBLP:conf/host/ChakrabortyMS17,kumar2017security}.
For example, as demonstrated in Fig.~\ref{fig:sca1}, a side channel attacker can monitor the timing information of the data movement into and out of the STT-MRAM memory array when the hardware runs cryptography applications.

To mitigate the security vulnerabilities in the STT-MRAM array, researchers have tried various strategies to wipe out the side channel signatures. Representative ones include parity encoding, short retention, and constant current write techniques \cite{DBLP:conf/dft/IyengarGRN16}.
To equalize the performance of different memory operations, these countermeasures need to slow down the fast operations and to increase the power consumption of low-energy operations. This brings significant design overheads in terms of delay and power. Next, as shown in Fig.~\ref{fig:sca2}, we demonstrate that \texttt{SpinCIM} is able to thwart side channel attacks on STT-MRAM memory array because of the following two features.

\begin{table*}[t]
\setlength{\tabcolsep}{14pt}
\small
\caption{{Key Parameters for MTJ Simulation.}}
\label{tab:parameter}
\centering
\begin{tabular}{ll}\toprule
 Parameter & Value \\ \midrule
 MTJ Surface Length & $40$ $nm$ \\
 MTJ Surface Width & $40$ $nm$ \\
 Spin Hall Angle & $0.3$ \\
 Resistance-Area Product of MTJ & $10^{-12}$ $\Omega \cdot m^2$ \\
 Oxide Barrier Thickness & $0.82$ $nm$ \\
 TMR & $100\%$ \\
 Saturation Field & $10^6$ $A/m$ \\
 Gilbert Damping Constant & $0.03$ \\
 Perpendicular Magnetic Anisotropy & $4.5 \times 10^5$ $A/m$ \\
 Temperature & $300 K$ \\
 \bottomrule
\end{tabular}
\end{table*}

\begin{table*}[t]
\setlength{\tabcolsep}{14pt}
\small
\caption{Delay and Energy Consumption of \texttt{Read/Write} Operations in Standard STT-MRAM Array.}
\label{tab:StandardTimingDelay}
\centering
\begin{tabular}{lrr} \toprule
 Operation & Delay ($ns$) & Energy ($fJ$) \\ \midrule
 Read `1' & 0.6 & 8.611 \\
 Read `0' & 0.6 & 7.669 \\
 Write `1' & 4.4 & 233.300 \\
 Write `0' & 3.3 & 191.400 \\
 \bottomrule
\end{tabular}
\end{table*}

\begin{table*}[tbp]
\setlength{\tabcolsep}{18pt}
\small
\caption{Delay and Energy Consumption of \texttt{Read/Write} and \texttt{SpinCIM} Operations in Enhanced STT-MRAM Array.}
\label{tab:SpinCIMTimingDelay}
\centering
\begin{tabular}{lrr} \toprule
Operation & Delay ($ns$) & Energy ($fJ$) \\ \midrule
Read `1' & 0.63 & 22.69 \\
Read `0' & 0.67 & 23.85 \\
Write `1' & 4.40 & 244.64 \\
Write `0' & 3.30 & 202.70 \\
CimNOT & 0.60 & 22.20 \\
CimAND & 0.55 & 22.30 \\
CimOR & 0.53 & 22.90 \\
CimNAND & 0.45 & 18.89 \\
CimNOR & 0.45 & 21.00 \\
CimXOR & 0.53 & 26.34 \\
CimADD & 0.53 & 26.32 \\
\bottomrule
\end{tabular}
\end{table*}

\noindent{\textbf{Increased operation types.}} In standard STT-MRAM array, there are only two operations between memory and CPU. The \texttt{Read} operation that reads data from specified memory address then sends the data to CPU; and the \texttt{Write} operation that receives data from CPU then writes the data into specified memory address.
As shown in Table~\ref{tab:StandardTimingDelay}, four levels of side channel information during reading `1',  reading `0',  writing `1',  and writing `0' can be monitored and analyzed. A side channel attacker will be able to differentiate the \texttt{Read/Write} operations, and analyze the number of `1'/`0' in a word with statistical methods.

However, in SpinCIM STT-MRAM array \cite{DBLP:journals/tvlsi/JainRRR18}, there are 11 possible operations. Besides the basic \texttt{Read} and \texttt{Write} operations, it also supports \texttt{CimNOT}, \texttt{CimAND}, \texttt{CimOR}, \texttt{CimNAND}, \texttt{CimNOR}, \texttt{CimXOR} and \texttt{CimADD} operations\footnote{Note that the number of enabled possible logic functions by \texttt{SpinCIM} depends on the peripheral circuits of the STT-MRAM array and it can slightly vary with different \texttt{SpinCIM} implementations.}.
Table~\ref{tab:SpinCIMTimingDelay} gives the delay and power of these SpinCIM operations. This makes side channel attacks much more complex. For instance, some of the logic computation operations such as \texttt{CimNOT}, \texttt{CimAND}, \texttt{CimOR}, \texttt{CimNAND}, \texttt{CimNOR}, \texttt{CimXOR} and \texttt{CimADD} have similar delay and power consumption as \texttt{Read} `1'/`0'. Therefore, identifying \texttt{Read} `1'/`0' will become more challenging. As another example, when SpinCIM logic computation results are required to be written into a specific memory address, which means a \texttt{SpinCIM} logic computation operation may be followed by a \texttt{Write} operation. In this case, a \texttt{SpinCIM} computation operation plus a \texttt{Write} `0' operation can obscure a \texttt{Write} `1' operation since they have similar delays and power consumption.

\noindent{\textbf{Reduced data transfers.}} In conventional logic computation instructions, the operation data needs to be read from the memory and sent to CPU, the computation result also needs to be sent back to the memory from CPU.
The operation data and the result data are transferred between CPU and memory though system bus, giving the side channel attacker more opportunities to exploit the information leakage in the system bus to reveal the transferred data.
While in \texttt{SpinCIM} logic computations, only the instructions that include the operation/result data addresses and the operation type are sent to the memory, then all the operations will be completed within the memory.
The operation mechanism of \texttt{SpinCIM} decreases the data transfers between CPU and memory, thus reducing the risks of being exploited and attacked by adversary side channel attackers.
{Note that there are various types of side-channel attacks. For those that exploit the operation type information, SpinCIM may confuse the attacker with the increased number of operation types. And for the cases that the attacker tries to utilize the operation data for secret analysis, \texttt{SpinCIM} prevents the attack from exploiting operation data information leakage in the system bus.}

{Current consumption trace matters in power-based side channel attacks, and the energy in Table~\ref{tab:StandardTimingDelay} and Table~\ref{tab:SpinCIMTimingDelay} is the integral of the product of current and voltage, which is closely related to the current consumption.
Assume that the objective of SCA is to retrieve the internal secret key $k$ of a crypto-algorithm, and the adversary can observe the input $p$ and the overall power consumption. The attacker will find an intermediate result $v$ (such as the current consumption trace) that depends on both $p$ and $k$. By observing the side-channel leakage of $v$, a hypothesis test on the key value $k$ can be created, and it can be expressed as: $L(k^*) = f_{k^*}(p) + \varepsilon$. The function $f_{k^*}$ is dependent on the crypto-algorithm and the specific implementation. The error $\varepsilon$ is an independent noise variable, defined by other unrelated activity in the crypto-implementation and measurement errors \cite{schaumont2012side}. Several types of power-based side channel analysis have been formulated starting from this relation, such as Correlation Power Analysis (CPA) and Simple Power Analysis (SPA). In both cases, for SpinCIM computing paradigm, the increased operation types could increase noise $\varepsilon$, and the reduced data transfer makes the exact formulation of function $f$ more difficult. As a result, SpinCIM offers considerable resilience against side channel attacks.}

\section{Security Issues in SpinCIM}\label{sec:cimsecurity}

Numerous advantages have been demonstrated in \texttt{SpinCIM}. From the performance perspective, it is able to alleviate the memory wall bottleneck in Von-Neumann computer structures. And from the security perspective, as demonstrated in this paper, it is a natural fit for some protective or preventive hardware security techniques.
However, the precondition to apply \texttt{SpinCIM} for performance improvement and security enhancement is that \texttt{SpinCIM} itself should be robust and secure enough.
In this section, we analyze the intrinsic security vulnerabilities in \texttt{SpinCIM}, and demonstrate a case study in practical attacking scenarios to gain a glimpse of the security issues within \texttt{SpinCIM}. Finally we discuss some potential mitigation techniques.

\subsection{SpinCIM Security Vulnerabilities}

\begin{figure}[t]
\centering
\subfigure[]{
\label{fig:SpinVulner1}
\includegraphics[width = 0.9\linewidth]{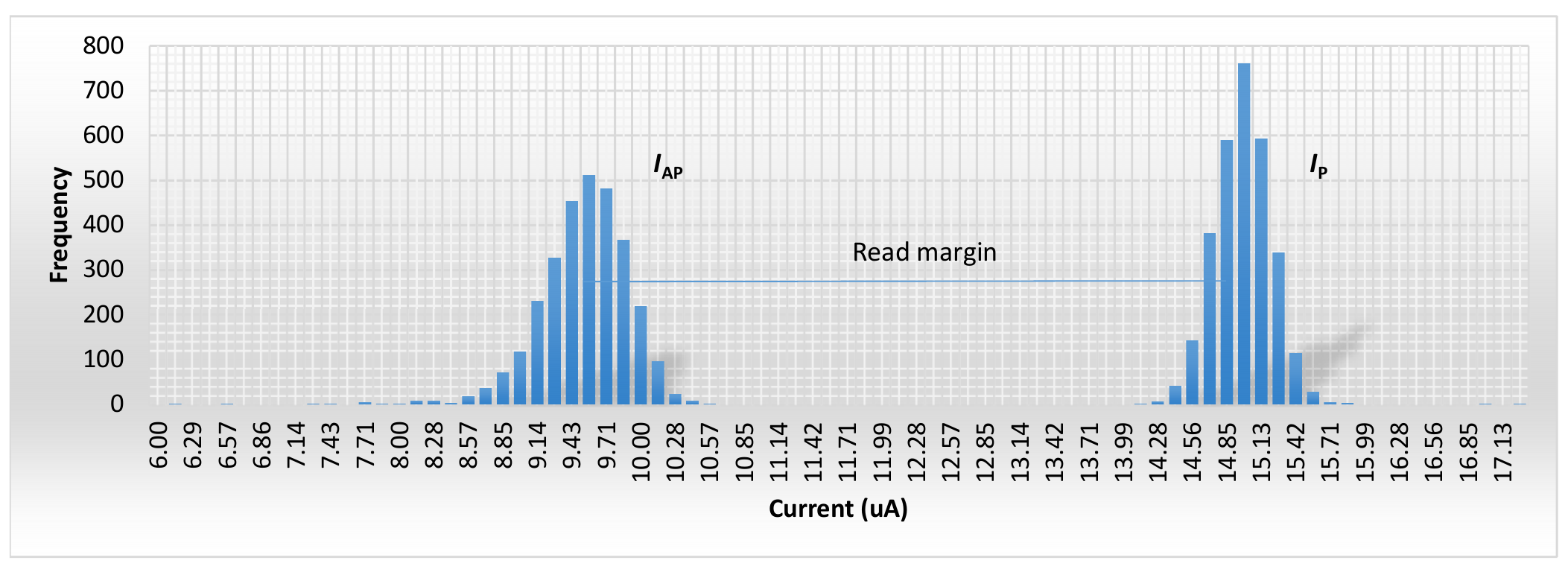}}
\subfigure[]{
\label{fig:SpinVulner2}
\includegraphics[width = 0.9\linewidth]{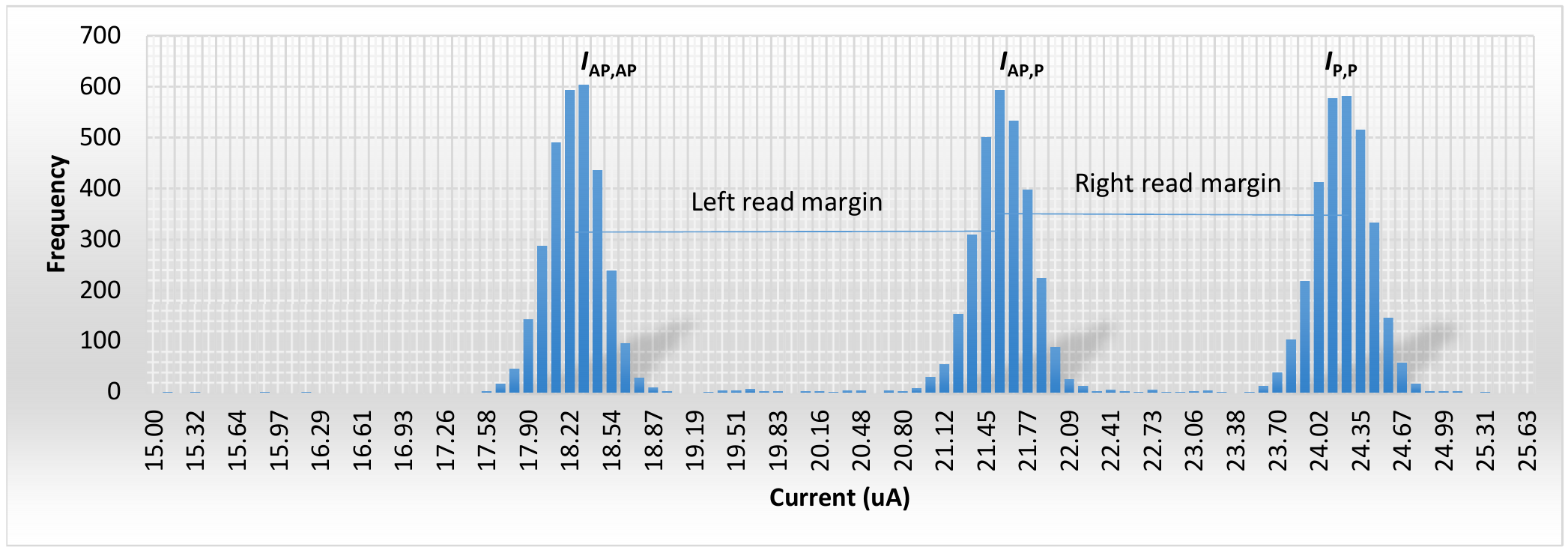}}
\caption{(a) Read margin in normal \texttt{Read} operation. (b) Read margins in \texttt{SpinCIM} operations.}
\end{figure}

In standard STT-MRAM array, the current flowing through SL ($I_{\rm SL}$) has two possible values, i.e. $I_{\rm P}$ and $I_{\rm AP}$. The TMR (($R_{\rm AP}-R_{\rm P}$)/$R_{\rm P}$) usually lies between 100\% and 200\% \cite{lu2015fully}. How to sense the current difference accurately to achieve reliable memory read has been a vital challenge for the commercial adoption of STT-MRAM. Both process variations and environmental factors (such as temperature and magnetic field) can affect the reliability of STT-MRAM. As a result, there have been extensive studies on this reliability issue and corresponding countermeasures, such as increasing TMR and adopting ECC (error-correction codes) strategies \cite{kang2015reconfigurable,vatankhahghadim2015variation,lee2010design,zhang2019spintronic,kwon2015high,del2014improving,kang2013low}. In this paper, we focus on the security and reliability issues that emerge with the new \texttt{SpinCIM} computing paradigm.

For \texttt{SpinCIM} operations, the reliability problem is more challenging.
Recall that in Fig.~\ref{fig:SpinCIM}, for two-input \texttt{SpinCIM} computations, the current flowing through SL is the sum of currents flowing through two MTJs, thus has three possible values, i.e. $I_{\rm P,P}$, $I_{\rm P,AP}$//$I_{\rm AP,P}$\footnote{This paper assumes $I_{\rm P,AP}$ equals to $I_{\rm AP,P}$, and the rest of the paper will denote both $I_{\rm P,AP}$ and $I_{\rm AP,P}$ with $I_{\rm AP,P}$.} and $I_{\rm AP,AP}$. Therefore, different with normal \texttt{Read} operation, which has only one read margin between $I_{\rm P}$ and $I_{\rm AP}$, in \texttt{SpinCIM} operations, there are two read margins, one between $I_{\rm AP,AP}$ and $I_{\rm AP,P}$, and the other one between $I_{\rm AP,P}$ and $I_{\rm P,P}$.
Simulation results in Fig.~\ref{fig:SpinVulner1} and Fig.~\ref{fig:SpinVulner2} demonstrate the read margin in conventional STT-MRAM array and the read margins in \texttt{SpinCIM}-enhanced STT-MRAM array.
We can see that each of the \texttt{SpinCIM} read margin is smaller than that in normal \texttt{Read} operation. Specifically, in normal \texttt{Read} operation, the read margin between $I_{\rm P}$ and $I_{\rm AP}$ is about $5.5$ $\mu A$, while in \texttt{SpinCIM} operations, the read margin between $I_{\rm AP,AP}$ and $I_{\rm AP,P}$ is about $3.2$ $\mu A$, and the read margin between $I_{\rm AP,P}$ and $I_{\rm P,P}$ is about $2.5$ $\mu A$.
The small read margins in \texttt{SpinCIM} could result in higher decision failure rates and make it more challenging to ensure reliable computations.
This can be exploited by malicious attackers and we will demonstrate a case study in practical scenarios.

\subsection{A Case Study in Authentication System}

In this part, we investigate the security vulnerabilities of \texttt{SpinCIM} with a case study of authentication system. It is demonstrated that an attacker is able to bypass the authentication by simply manipulating the thermal conditions of STT-MRAM array.

Authentication system is widely used in most of the information systems and services.
It provides access control for protected resources (such as the credit card numbers) by checking whether a user's credentials match the authorized users database.
In normal modes, users are identified with a username and a password. When the username and password are both correct, the user is authorized to access the system. Such access control principle can be implemented with the following high-level programming sentences:

\vspace{3mm}
\fbox{%
  \parbox{0.95\textwidth}{%
    \begin{center}
      \textsc{IF (username is correct {AND} password is correct)}\\
      \textsc{\{THEN validate pass, enter the system!}\}
    \end{center}
  }%
}
\vspace{3mm}

In the low-level implementation, it needs to check the username and password that are typed in by the user (represented as $\rm {u_{t}}$ and $\rm {p_{t}}$) with the username and password items in the authorized users database (represented as $\rm {u_{d}}$ and $\rm {p_{d}}$). {Checking whether two items are identical can be achieved with the \texttt{XNOR} operation: $x$ \texttt{XNOR} $y$ = 1 when $x$ equals $y$, otherwise, $x$ \texttt{XNOR} $y$ = 0.}
To reduce the data transfer bottleneck, the authentication process may be accomplished in the enhanced STT-MRAM array.
Thus the user is authorized to the system \textbf{if and only if}:
\begin{equation}\label{equ:judgment}
(\rm {u_{t}}~\texttt{CimXNOR}~\rm {u_{d}})~\texttt{CimAND}~(\rm {p_{t}}~\texttt{CimXNOR}~\rm {p_{d}}) = 1
\end{equation}
\texttt{CimXNOR} operation can be typically realized as below \cite{DBLP:journals/tvlsi/JainRRR18}:
\begin{equation}\label{equ:xor1}
\rm {u_{t}}~\texttt{CimXNOR}~\rm {u_{d}} = (\rm {u_{t}} ~\texttt{CimAND} ~\rm {u_{d}}) ~\texttt{OR} ~(\rm {u_{t}} ~\texttt{CimNOR} ~\rm {u_{d}})
\end{equation}
\begin{equation}\label{equ:xor2}
\rm {p_{t}}~\texttt{CimXNOR}~ \rm {p_{d}} = (\rm {p_{t}} ~\texttt{CimAND} ~\rm {p_{d}}) ~\texttt{OR} ~(\rm {p_{t}} ~\texttt{CimNOR} ~\rm {p_{d}})
\end{equation}

Assume that a malicious attacker judiciously trigger a certain mistake to perform certain \texttt{CimAND} function as \texttt{CimOR} (the feasibility will be discussed later), then the attacker is able to conduct effective bypass access control attack.


By triggering the \texttt{CimAND} in Equation (\ref{equ:judgment}) as \texttt{CimOR}, the judgment becomes:
\begin{equation*}
(\rm {u_{t}}~\texttt{CimXNOR}~\rm {u_{d}})~\texttt{CimOR}~(\rm {p_{t}}~\texttt{CimXNOR}~\rm {p_{d}})
\end{equation*}

Therefore, the attacker gains access to the system when either the username is correct or the password is correct. The corresponding high-level control sentence is:

\vspace{3mm}
\fbox{%
  \parbox{0.95\textwidth}{%
    \begin{center}
      \textsc{IF (username is correct {OR} password is correct)}\\
      \textsc{\{THEN validate pass, enter the system!}\}
    \end{center}
  }%
}
\vspace{3mm}

As a result, even though the attacker is not aware of the passwords, he is able to access the system simply with a correct username. And username is easy to be pirated because it is mostly related to the user's phone numbers or real names thus easy to guess, and usually not technically protected.

A more powerful attack can be launched to bypass both user and password authentications.
By triggering \texttt{CimAND} in Equation~(\ref{equ:xor1}) as \texttt{CimOR}, it becomes
\begin{align*}
&(\rm {u_{t}} ~\texttt{CimOR} ~\rm {u_{d}}) ~\texttt{OR} ~(\rm {u_{t}} ~\texttt{CimNOR} ~\rm {u_{d}})\\
&=(\rm {u_{t}}+\rm {u_{d}})+\overline{\rm {u_{t}}+\rm {u_{d}}}\\
&=1
\end{align*}
Similar manipulation can be applied to Equation~(\ref{equ:xor2}).
Then Equation~(\ref{equ:judgment}) becomes
\begin{equation*}
1~\texttt{CimAND}~1 = 1.
\end{equation*}

Corresponding high-level access control sentence is

 \vspace{3mm}
\fbox{%
  \parbox{0.95\textwidth}{%
    \begin{center}
      \textsc{IF (Always True)}\\
      \textsc{\{THEN validate pass, enter the system!}\}
    \end{center}
  }%
}
\vspace{3mm}

As a result, all users that attempt to gain access to the system, including the illegal users, are authorized to enter the system.

\begin{figure}[t]
\centering
\includegraphics[width = 0.75\linewidth]{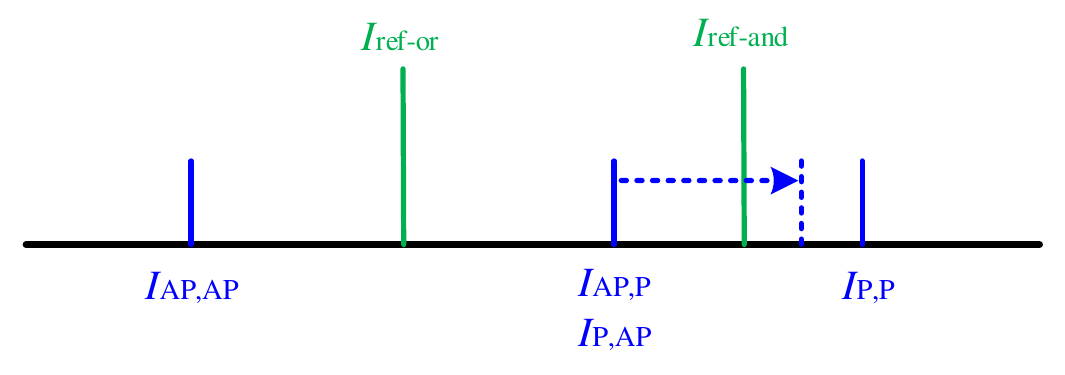}
\caption{Confusion between $I_{\rm AP,P}$ and $I_{\rm ref-and}$.}
\label{fig:ref}
\end{figure}

\noindent{\textbf{Feasibility of triggering \texttt{CimAND} as \texttt{CimOR}}}.
Table~\ref{tab:CimAndOrCOM} demonstrates the truth table of \texttt{CimAND} and \texttt{CimOR} operations. Only when the two MTJ states are \texttt{AP} and \texttt{P}, \texttt{CimAND} and \texttt{CimOR} have different outputs: logic `0' and logic `1', respectively. Therefore, as demonstrated in Fig.~\ref{fig:ref}, to trick \texttt{CimAND} into \texttt{CimOR}, $I_{\rm AP,P}$ needs to be sensed as larger than $I_{\rm ref-and}$.
Recall that the read margin between $I_{\rm AP,P}$ and $I_{\rm P,P}$ is smaller than other read margins, it is easier to confuse $I_{\rm AP,P}$ with $I_{\rm ref-and}$.
In this case, the output under \texttt{AP}, \texttt{P} will be mistakenly computed as logic `1', which is the correct output of \texttt{CimOR}.

\begin{table*}
\setlength{\tabcolsep}{13pt}
\small
\begin{threeparttable}
\caption{Truth Table for \texttt{CimAND} and \texttt{CimOR} Operations.}
\label{tab:CimAndOrCOM}
\centering
\begin{tabular}{ccccc} \toprule
 \multirow{2}{*}{Logic} & \multirow{2}{*}{MTJ States} & \multirow{2}{*}{$I_{\rm SL}$} & \multicolumn{2}{c}{Function} \\ \cline{4-5}
  & & & {CimAND}\tnote{\dag} & {CimOR}\tnote{\ddag} \\ \midrule
 0, 0 & AP, AP & $I_{\rm AP,AP}$ & 0 & 0 \\
 0, 1 & AP, P & $I_{\rm AP,P}$ & 0 & 1 \\
 1, 0 & P, AP & $I_{\rm AP,P}$ & 0 & 1 \\
 1, 1 & P, P & $I_{\rm P,P}$ & 1 & 1 \\
 \bottomrule
\end{tabular}
\begin{tablenotes}
	\item[\dag] $I_{\rm ref-and}\in (I_{\rm AP,P},I_{\rm P,P})$
	\item[\ddag] $I_{\rm ref-or}\in (I_{\rm AP,AP},I_{\rm AP,P})$
\end{tablenotes}
\end{threeparttable}
\end{table*}

\vspace{3mm}

\begin{table*}[htbp]
\setlength{\extrarowheight}{2.0pt}
\small
\caption{\texttt{CimAND} Failure Rates Under Natural Conditions and  Laser Heat Conditions.}
\label{tab:CimAndFailure}
\centering
\begin{tabular}{c|c|c|c}
 \specialrule{0.8pt}{0pt}{0pt}
 \multirow{2}{*}{Failure} & Natural & \multicolumn{2}{c}{With Laser Heat} \\
 \cline{2-2} \cline{3-4}
 & $20$\textcelsius & $50$\textcelsius & $100$\textcelsius \\ \hline
 $I_{\rm AP,P}$ > $I_{\rm ref-and}$ & $0.5\%$ & $0.6\%$ & $4.4\%$ \\
 $I_{\rm AP,AP}$ > $I_{\rm ref-and}$ & $0$ & $0$ & $0.3\%$ \\
 \specialrule{0.8pt}{0pt}{0pt}
\end{tabular}
\end{table*}


To evaluate the failure rate of performing \texttt{CimAND} as \texttt{CimOR} operation, we conduct 10000 Monte Carlo simulations for \texttt{CimAND} operations under natural conditions and intentional attack conditions.
{In the intentional attacking scenario, the target MTJs are dealt with laser heat in the same way as \cite{wu2016temperature}. }
While the resistance of \texttt{P}-state MTJ is relatively stable under laser heat conditions, the resistance of \texttt{AP}-state MTJ demonstrates an obvious declining trend. Accordingly, under laser heat, $I_{\rm AP,P}$ will slightly increase, and be more closer to $I_{\rm ref-and}$, resulting in a higher possibility to sense $I_{\rm AP,P}$ as larger than $I_{\rm ref-and}$.
The simulation results in Table~\ref{tab:CimAndFailure} validate that the main failure of \texttt{CimAND} come from $I_{\rm AP,P}$ > $I_{\rm ref-and}$. Under natural conditions, the failure rate of $I_{\rm AP,P}$ > $I_{\rm ref-and}$ is $0.5\%$.
After intentional laser heat to $100$\textcelsius, the failure rate of $I_{\rm AP,P}$ > $I_{\rm ref-and}$ increases to $4.4\%$, which is $8.8\times$ higher than the natural conditions. Such high failure rate of \texttt{CimAND} greatly threatens the security of computer systems that employ \texttt{SpinCIM}.


\subsection{SpinCIM Security Vulnerabilities Mitigation}

To relieve the reliability problem in \texttt{SpinCIM} computations, extending ECC strategies \cite{DBLP:journals/tvlsi/JainRRR18} have been proposed.
It tries to detect and correct the sensing errors. However, as demonstrated in Fig.~\ref{fig:SpinCIM}, \texttt{SpinCIM} does not sense the individual resistance state of the input MTJs. Instead, only the sum of the currents of the two MTJs matters.
Therefore, the extended ECC scheme to correct the sense errors in input MTJs is insufficient in correcting errors in \texttt{SpinCIM} computations.

Recall that the factors that affect the robustness of \texttt{SpinCIM} are the degraded sensing margins and external disturbance.
Increasing the TMR of MTJ to increase the sensing margins is one effective way, however, it is out of the scope of this paper and much research has put emphasis on this topic.
For the external disturbance, a possible countermeasure is to design sense amplifier in a disturbance-aware way. Take the laser heat disturbance for example, assuming that before laser heated, the three possible MTJ currents are $I_{\rm AP,AP}$, $I_{\rm AP,P}$ and $I_{\rm P,P}$, and after being laser heated, they become $\left( I_{\rm AP,AP}+\alpha \right)$, $\left( I_{\rm AP,P}+\beta \right)$ and $\left(I_{\rm P,P}+\gamma \right)$, respectively, where $0<\alpha<\beta<\gamma$.
Then the reference currents of \texttt{CimOR}, namely $I_{\rm ref-or}$, can be accordingly adjusted from $\frac{I_{\rm AP,AP}+I_{\rm AP,P}}{2}$ to $\frac{I_{\rm AP,AP}+I_{\rm AP,P}+\alpha+\beta}{2}$, and $I_{\rm ref-and}$ can be accordingly adjusted from $\frac{I_{\rm AP,P}+I_{\rm P,P}}{2}$ to $\frac{I_{\rm AP,P}+I_{\rm P,P}+\beta+\gamma}{2}$.
{Although a preliminary thought of the possible mitigation method is discussed, its implementation detail and its effectiveness need much more further in-depth study. In addition, how to assure the reliability of SpinCIM remains an open question, we may focus on this issue in the future work.}

\section{Outlook for SpinCIM Security}\label{sec:outlook}



With the unique features of computing, \texttt{SpinCIM} has demonstrated a lot of advantages in facilitating hardware security techniques. For circuit obfuscation techniques, functions that are implemented in enhanced STT-MRAM array can be judiciously hidden from a reverse engineering attacker. As for thwarting side channel attack, \texttt{SpinCIM} reduces the data transfers between CPU and memory thus reduces side channel information leakages, and increases the operation types within the STT-MRAM array, thus complicates the malicious power or timing side channel analysis. 
However, in addition to these positive features, \texttt{SpinCIM} also exposes some security vulnerabilities due to the degraded read margins and being sensitive to external disturbance.
A case study in authentication systems demonstrates that an attacker is able to achieve Trojan-similar attacks by simply manipulating the thermal conditions of STT-MRAM, even without the need for circuit modifications. Possible mitigation methods are discussed while future in-depth study is still needed.

Perhaps the biggest challenge for both \texttt{SpinCIM} and \texttt{SpinCIM} security applications is the code mapping and the data mapping problem. 
Code mapping decides which operations should be executed in memory and which operations should be executed in CPU and how to make them co-operate efficiently, while data mapping decides how should data be mapped to the STT-MRAM array.
Most of the CIM techniques only support computations when the operation data are stored in the same bank, mapped to different rows, and stored in the same set of columns.
Code mapping and data mapping research need collaboration work from the device, logic, architecture and application levels. Only when the \texttt{SpinCIM} computing paradigm becomes mature in all design levels, its application for hardware security can be enriched, and its intrinsic security vulnerabilities can be solved better.

Also previously, security has often been considered as an afterthought, with performance dominating the design requirements, resulting in numerous security vulnerabilities. 
While \texttt{SpinCIM} is newly proposed and still in exploration progress, it provides an opportunity to reconsider security as a first class requirement at the design stage. 
Research on \texttt{SpinCIM} security should synchronize with the study in \texttt{SpinCIM} computing paradigm. 
To this end, more hardware security opportunities and challenges that are related to \texttt{SpinCIM} will emerge with the more and more in-depth investigation in \texttt{SpinCIM}, thus deserves to be given full attention in future works.

\section{Conclusions}\label{sec:conclu}

Spin-based computing in memory techniques have demonstrated promise in alleviating memory wall challenges in traditional Von-Neumann architectures, thus has attracted attention from both industry and academia communities.
In this paper, we have studied \texttt{SpinCIM} from a security perspective. 
We have investigated the feasibility to enhance hardware security with the unique properties in \texttt{SpinCIM} computing paradigm and found that \texttt{SpinCIM} was a natural fit for some security applications.
We have also discussed about the possible security vulnerabilities in \texttt{SpinCIM} and demonstrated with a case study in practical attacking scenarios, then discussed possible mitigation techniques and presented an outlook to the future research for \texttt{SpinCIM} security.


\bibliographystyle{unsrt}

\end{document}